\documentclass[reprint,
superscriptaddress,
amsmath,amssymb,aps,prb,]{revtex4-2}
\usepackage{graphicx}
\usepackage{amssymb}
\usepackage{amsthm}
\usepackage{hyperref}
\usepackage{color}
\usepackage{xcolor}
\usepackage{lineno}
\modulolinenumbers[1]

\begin{document}

\title{Temperature effects on the point defects formation in [111] W by neutron 
induced collision cascade}

\author{F. J. Dom\'inguez-Guti\'errez}
\email{Corresponding author: javier.dominguez@ncbj.gov.pl}
\affiliation{NOMATEN Centre of Excellence, National Centre for Nuclear 
Research, ul. A. Sołtana 7, 05-400 Swierk/Otwock, Poland}
\affiliation{Institute for Advanced Computational Science, Stony Brook University, Stony Brook, NY 11749, USA}

\begin{abstract}
Tungsten is used as plasma-facing wall in ITER where it is subjected to extreme operating 
conditions. 
In this work, we study the damage formation in [111] crystalline W by neutron bombardment 
in the temperature range of 300-900 K which is important for designing  next generation 
of fusion reactors. 
The Molecular Dynamics (MD) simulations are performed at a primary knock-on atoms (PKA) 
energy of 1 keV within the Gaussian Approximation Potential (GAP) framework. 
The analysis of the induced damage is done by the Fingerprinting and Visualization Analyzer
of Defects (FaVAD) which is based on a rotation invariant mapping of the local 
atomic neighborhoods allowing to extract, to identify and to visualize the atomic defect 
formation dynamically during and after the collision cascades. 
The evaluation allows to classify the various defect types and to quantify the defect evolution
as function of time and initial sample temperature. 
We found that the production of Frenkel pair increases as a function 
of the temperature due to thermal activated mechanisms.

\end{abstract}


\maketitle
\section{Introduction}
\label{sec:intro}

 Tungsten is considered as a Plasma Facing Material (PFM) due to its physical and 
 chemical properties such as low erosion rates, small tritium retention, and high 
 melting point which is important to design the next generation of fusion machines. 
 Mechanical properties of W can be enhanced or modified due to irradiation where formed
permanent point or extended defects (dislocations) can increase the hardness of the material 
\cite{wirth_hu_kohnert_xu_2015,Mason_2018,FIKAR201860}. 
 For this reason, experimental exploration of the characterization of PFMs needs to be guided
 by  numerical modeling saving laboratory and financial resources for carrying out
 experiments at extreme environments \cite{EHRLICH200079,Bonny_2014,Herrmann_nucl_fusion,BOLT200243}, 
 where atomistic simulations method can be performed 
 \cite{Marian_2017,Jav_UvT,NORDLUND1995448,Nor05c,Jesper_GAP}.

In order to model the formation of point defects due to neutron bombardment, 
the Molecular dynamics (MD) method is applied to perform simulations of collision 
cascades. The accuracy of the numerical results is highly dependent on the 
interatomic potentials capable to describe induced damage in materials \cite{Nor18}.
However, traditional potentials based on the embedded atom model (EAM) are limited 
to functional forms \cite{PhysRevB.29.6443,doi:10.1063/1.2336465} and are in risk to
wrongly model some point defects lacking of physical meaning in material
damaging processes \cite{DOMINGUEZGUTIERREZ2020100724}.
Thus, interatomic potentials developed by using machine learning (ML) techniques  
are recently used to perform MD simulations with an accuracy close to 
Density Functional Theory (DFT) 
\cite{Jesper_GAP,PhysRevLett.104.136403,doi:10.1080/21663831.2020.1771451} and
systematically improved towards the accuracy of the training data set. 
In our previous work, we applied interatomic potentials based on the Gaussian Approximation 
Potential (GAP) framework to numerically model the damage in crystalline materials 
due to irradiation in a fusion reactor \cite{DOMINGUEZGUTIERREZ201756,Jesper_GAP}.  
MD simulations were analyzed by a recently developed workflow for semi-automatic 
identification and classification of defects in crystalline structures and reported 
results are in good agreement with experimental data \cite{VONTOUSSAINT2021107816}.
Therefore, the goal of the present work is to better understand the thermal activated 
mechanisms for material damage in crystalline W samples which is commonly used in experiments 
for irradiation in a fusion reactor. 
For this, we perform MD simulations to model neutron irradiation in [111] W at 1 keV of PKA in 
a sample temperature range from 300 to 900 K providing an analysis of the point defects 
formation after collision cascades.

Our paper is organized as follows: in Section \ref{sec:methods} 
we briefly discuss the theory to develop the machine learned (ML) potential 
\cite{2006.14365,Jesper_GAP} into the GAP framework, as well as the software workflow for 
the Fingerprinting and Visualization Analyzer of Defects (FaVAD) \cite{VONTOUSSAINT2021107816,favad}. 
Our results for the total number of points defects and atomic displacement as a function 
of the sample temperature are presented in Sec. \ref{sec:results}. 
Finally, in section \ref{conclusions}, we provide concluding remarks.

\section{Methods}
\label{sec:methods}

\begin{figure}[b!]
   \centering
   \includegraphics[width=0.35\textwidth]{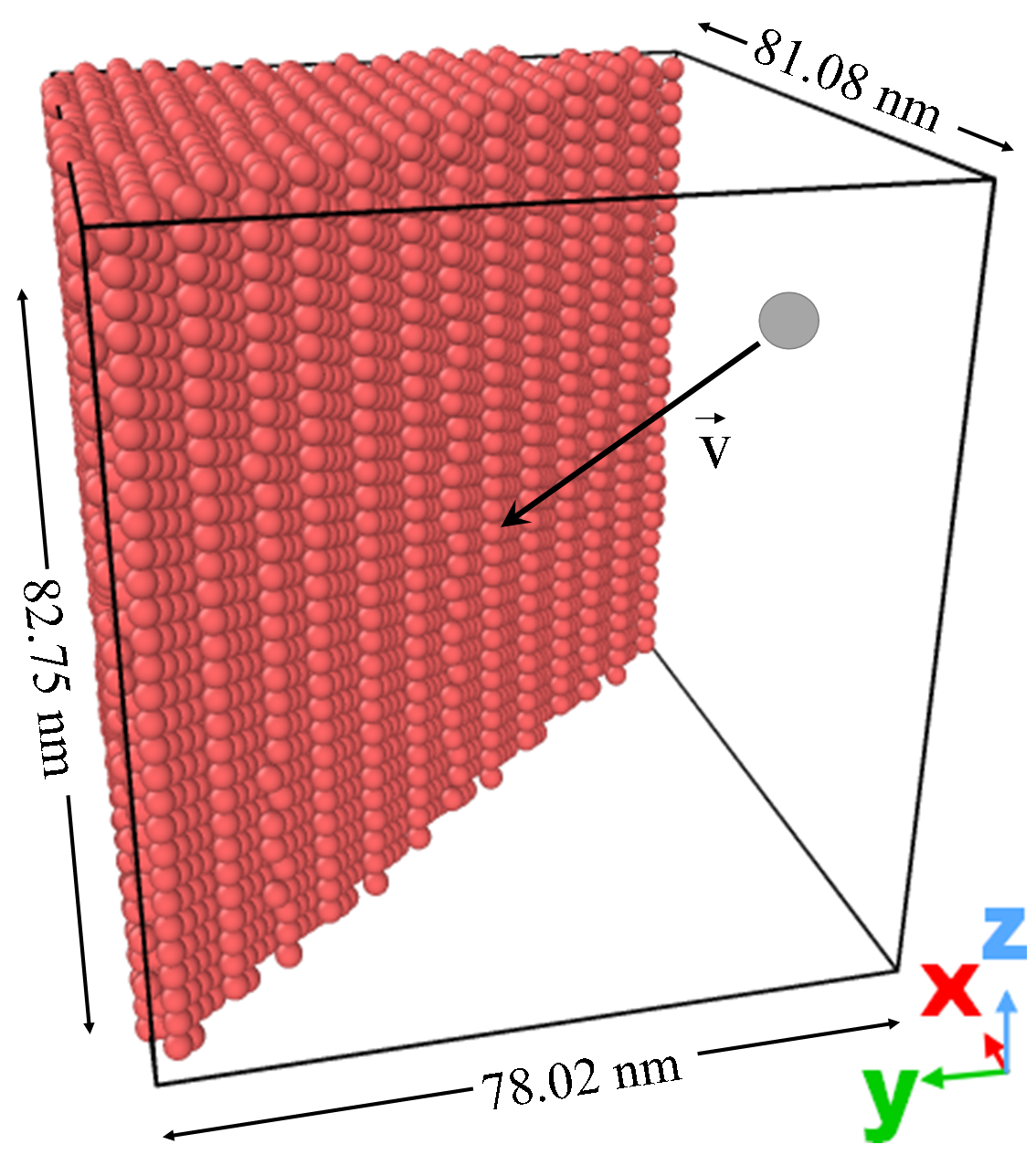}
   \caption{(Color on-line). Schematic of the prepared numerical box used in the 
   MD simulations with GAP potentials. 
   The sample was sliced on the [-110] $\times$ [-1-12] plane for better visualization of 
   the crystal orientation and projectile's initial position into the sample, as well as the 
   orientation of its velocity vector.}
   \label{fig:fig1}
\end{figure}

MD simulations are traditionally used to model neutron bombardment processes at 
different PKA energies 
providing information of damage in crystalline materials.   
In our work, we start by defining a numerical box with a size of 
(81.08 $\hat x$, 78.02 $\hat y$, 82.75 $\hat z$) nm and W atoms arranged 
into a crystalline body-centered-cubic (BCC) geometry with a lattice 
constant of $a = 3.18$ \AA{}; according to DFT calculations for computing the
GAP potentials \cite{Jesper_GAP}.
Then, the numerical sample is prepared by a process of energy optimization
and thermalization using the Langevin thermostat, 
with the time constant of 100 fs. \cite{DOMINGUEZGUTIERREZ201756}. 

In radiation experiments, the energy of neutrons is around 14 MeV after
D+T reaction. 
Particles can lose energy through non-ionizing interactions with
the materials through displacement damage where the collision energy 
between incoming particle and a lattice nucleus (Primary Knock-on 
Atom) is in the order of 10$^{3} -$ 10$^{5}$ eV; displacing atoms 
from original lattice positions to generate point defects.
Thus, it has been reported that 1 keV PKA value is at the subcascade 
energy threshold to produce more than one stable Frenkel pair in 
crystalline W samples at room temperature \cite{Nor17b}. 
This kinetic energy is assigned to a W atom located at ten different 
randomly chosen locations with a velocity 
vector oriented to the center of the sample, as shown in 
Fig. \ref{fig:fig1} to mimic damage production at the experimental 
depths of hundreds of nanometers. 
The W samples are equilibrated to temperatures of 400, 600, 800, and
900 K to consider conditions into nuclear reactors. 
The room temperature is used in our work as a reference 
\cite{doi:10.1063/1.2336465}.
Then, the Velocity Verlet integration algorithm is applied to model the 
collision cascade for 6 ps, followed by an additional relaxation time of 4 ps. 
The MD simulations were performed in the institutional cluster of the Stony Brook University 
by using the 
Large-scale Atomic/Molecular Massively Parallel Simulator (LAMMPS) 
\cite{PLIMPTON19951}. 
\subsection{Gaussian Approximation Potential framework}
\label{sec:modified_GAP}

The Quantum mechanics and 
Interatomic Potential package (QUIP) \cite{quip} is used as an interface to implement machine learned interatomic potentials
based on GAP \cite{PhysRevLett.104.136403} that can be 
systematically improved by their training data set modifications. 
The GAP potential models collision cascades mechanisms due to its training data set for 
describing short-range dynamics, the liquid phase, and the re-crystallization process 
where defect creation and annihilation tend to happen 
\cite{Jesper_GAP,DOMINGUEZGUTIERREZ2020100724}.
Here, the total energy of a system of $N$ atoms is expressed into the GAP framework as 
\begin{equation}
    E_{\textrm{tot}} = \sum^N_{i<j} V_{\textrm{pair}}(r_{ij}) + 
    \sum^{N_d}_{i} E^i_{\textrm{GAP}},
\end{equation}
where $V_\mathrm{pair}$ is a purely repulsive screened Coulomb
potential.
The term $ E_{\textrm{GAP}}$ is obtained by using the Smooth Overlap of Atomic Positions 
(SOAP) package \cite{PhysRevLett.104.136403} with $N_d$ as the number of 
descriptor environments (DE), $ \vec{\xi}^{\ i}$, for the $N$-atom system. 
Which describes most of the interatomic bond energies by a many-body term 
\cite{Jesper_GAP,2006.14365} that are obtained from the atomic density 
$\rho^{{} i}$ between the atoms $i$ and $j$ as \cite{PhysRevB.87.184115},

\begin{eqnarray}
\rho^{{} i}(\vec r) 
     & = & \sum_{nlm}^{NLM} c^{(i)}_{nlm}g_n(r)Y_{lm}\left(\hat r\right),
\label{eq:Eq2}
\end{eqnarray}
with $Y_{lm}(\hat r)$ as the spherical harmonics and a set of basis functions 
in radial directions $g_n(r)$ as $c^{(i)}_{nlm} = \langle g_n Y_{lm} | \rho^i \rangle$ 
\cite{PhysRevB.87.184115,Jav_UvT}. 
Thus, the sum over the order $m$ of the squared modulus of the coefficients $c_{nlm}$ that
is invariant under rotations around the central atom \cite{PhysRevB.90.104108} defines the 
DE as
\begin{equation}
\vec{\xi}_{k}^{\ i} = \left\{ \sum_m
\left(c_{nlm}^i \right)^* c_{n'lm}^i \right\}_{\ n,n',l},
\label{eq:Eq3}
\end{equation}
where $c^{*}_{nlm}$ denotes the complex conjugate of $c_{nlm}$.
In order to perform MD simulations for collisions cascades, t
he original GAP for W was modified to include a realistic repulsive 
pair potential, $V_{\rm pair}$, according to methodology used
to develop EAM and Tersoff-like potentials.
Here, the smooth connection between the trained GAP and 
$V_{\rm pair}$ is done by adding dimer distances larger than 
1.1 \AA{} in the training database.
In addition, several short-range environments are included
in the training data set by randomly placing atoms to both 
perfect crystalline structures and systems 
containing one or two vacancies; capturing the short-range 
many-body dynamics in collision cascades for BCC W samples.
Finally, it is important to note that the modified GAP becomes 
numerically unstable at internuclear distance below 0.03 \AA{} 
which cannot be reached in high-energy cascade simulations.

\subsection{Fingerprinting and Visualization Analyzer of Defects}
\label{sec:DV_method}

The machine learning based software workflow FaVAD
\cite{VONTOUSSAINT2021107816,favad} is applied to the output data of the MD simulations 
to identify and quantify the number of formed self interstitial atoms (SIAs) and vacancies 
during collision cascades.
FaVAD describes the local atomic environment of the $i$-th atom of the damaged 
material by the normalized DE $\vec{q}^{\ i} = \vec{\xi}^{\ i}/|\vec{\xi}^{\ i}|$ 
and compares to those for a defect free environment, at a given temperature, to identify 
the point defects in the damaged material by computing their distance difference as
\cite{Jav_UvT,VONTOUSSAINT2021107816}.
\begin{equation}
    d^M (T) =
    \sqrt{ \left( \vec{q}^{\ i} - \vec{Q}\left(T\right) \right)^{\textrm{T}} 
\Sigma^{-1} (T) \left( \vec{q}^{\ i} - \vec{Q}\left(T\right) \right)},
\label{eq:maha}
\end{equation} 
where $\vec{Q}\left(T\right)=\frac{1}{N}\sum_{i=1}^{N}\vec{q}^{i}\left(T\right)$ is 
the mean of the DEs of the defect free sample; and $\Sigma$ is the associated 
co-variance matrix of the DE components \cite{Maha,Jav_UvT}.
This calculation allows us to quantify point defects at different temperatures, 
where thermal motion is very important. 
The identification of vacancies is done by the computation of the nearest neighbor distance
between the position of the damaged sample atoms and sampling grid points that defines 
the spatial volume of the identified vacancy \cite{2004.08184,VONTOUSSAINT2021107816}.


\section{Results and Discussion}
\label{sec:results}

\begin{figure}[b!]
   \centering
   \includegraphics[width=0.45\textwidth]{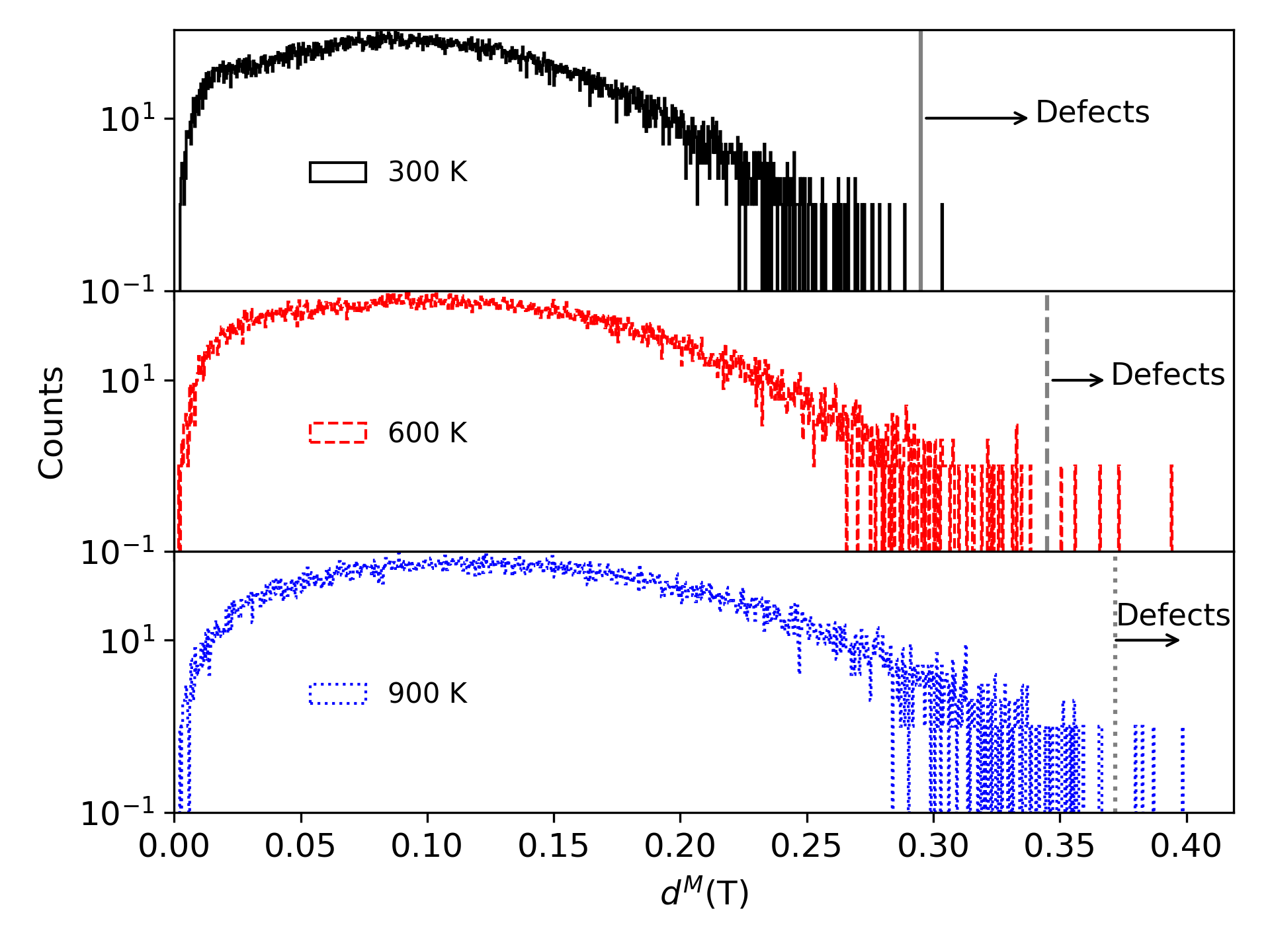}
   \caption{(Color on-line). Distance difference, $d^{M}(T)$, between the defect free 
   environment and the W damaged sample at different temperatures. 
   A threshold is defined for each case to identify and quantify points defects 
   by considering the atomic thermal motion.}
   \label{fig:fig2}
\end{figure}

The analysis of the final frame of the MD simulations provides the information of the 
point defects formed that can be identified and quantified by FaVAD. 
In Fig. \ref{fig:fig2}, we present results for the comparison between 
the damage material environment to a defect free one, for different 
sample temperatures.
For the chosen MD simulations, the velocity of the projectile is parallel 
to the [111] orientation.
Note that the thermal motion due to height temperature is observed by 
the lattice distortion at 0.25 to 0.35 $d^{M}(T)$. 
The threshold that defines the actual self interstitial atoms (SIAs) 
in the sample needs to be assigned according to the sample temperature, 
where a gap between these defects and atoms in their lattice position is observed 
for all the cases.
FaVAD identified and quantified the production of Frenkel pairs for all the sample where 
the maximum number of defects is found at 600 K, and decreases at higher temperatures. 
At a PKA of 1 keV and due to the training data set of the GAP potentials, FaVAD did not find 
another type of defect. 
In Tab. \ref{tab:MDdata}, we report the number of identified and quantified point defects 
in the damaged material, as reference, and noticing an increase of the Frenkel pair 
production due to high sample temperature.

\begin{table}[b!]
    \centering
    \caption{Average number of point defects and vacancies as a 
    function of the temperature (in Kelvin) identified by FaVAD. 
    SIA are identified as W atoms with the highest probability 
    to be in an interstitial site.}
    \begin{tabular}{l r r r r r}
       \hline
        \textbf{Defect} & 300 & 400 & 600 & 800 & 900  \\
        \hline
        SIA         & 2 $\pm$ 1   & 3 $\pm$1     & 5 $\pm$ 2   & 4$\pm$1 & 4 $\pm$ 1 \\
        NtV.        & 14 $\pm$ 2  & 18 $\pm$ 2   & 33 $\pm$ 5  & 27 $\pm$ 3 & 25 $\pm$ 3 \\
        Total       & 16 $\pm$ 2  & 21 $\pm$2 2  & 38 $\pm$ 5  & 31 $\pm$ 3 & 29 $\pm$ 2 \\
        \hline
        Vac.         & 2 $\pm$ 1   & 3 $\pm$1     & 5 $\pm$ 2   & 4$\pm$1 & 4 $\pm$ 1 \\
        \hline
    \end{tabular}
    \label{tab:MDdata}
\end{table}

\begin{figure}[t!]
   \centering
   \includegraphics[width=0.45\textwidth]{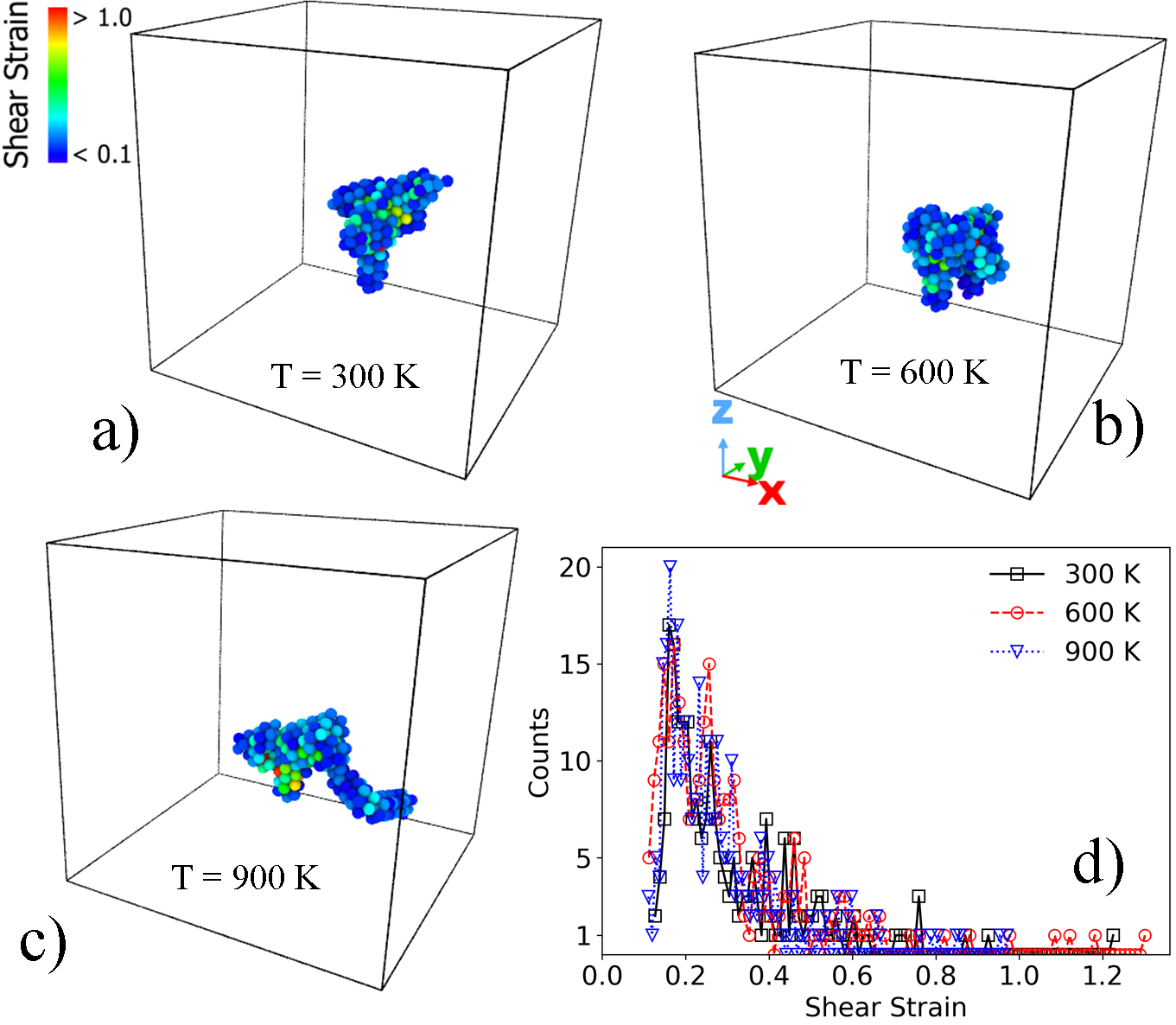}
   \caption{(Color on-line). von Mises strain distribution for the identified point defects at 
   a sample temperature of 300K in a), 600 K in b), and 900 K in c). 
   A comparison of the results is presented in Fig. d) as a histogram.}
   \label{fig:fig3}
\end{figure}

In order to investigate the sample temperature effects on the material's damage. 
We compute the von Mises atomic strain by considering the atomic 
distance difference, 
$\textbf{d}$, between the the $m$-th nearest neighbors of 
the $n$-th atom of the pristine and damaged samples as:
\begin{equation}
\boldsymbol{J}_{n} =  \left(\sum_{m}\boldsymbol{d}_{m}^{0T}\boldsymbol{d}_{m}^{0} 
    \right)^{-1} 
    \left(\sum_{m}\boldsymbol{d}_{m}^{0T}
    \boldsymbol{d}_{m} \right).
\end{equation}
Thus, defining the Lagrangian strain matrix of the $n$-th atom as
\cite{2007MJ200769}:
\begin{equation}
    \boldsymbol{\eta}_n =  1/2 \left(\boldsymbol{J}_{n} \boldsymbol{J}_{n}^T-I\right),
\end{equation}
and the corresponding von Mises strain of the $n$-th atom is computed as:
\begin{equation}
    \eta_n =   \sqrt{\frac{\zeta_{ij}\zeta_{ij}}{2}}, \quad 
    {\rm with} \quad
    \zeta_{ij} =  \eta_{ij}-\eta_{kk} \delta_{ij}. 
\end{equation}
This approach is implemented in OVITO \cite{ovito}. 
In Fig. 3, we report the von Mises strain distribution of the damaged sample 
at 300 in a), 600 in b), and 900 K in c). 
Fig. \ref{fig:fig3} d) shows a histogram of the results comparison, as a reference.
Atoms with strain values less than 0.1 are associated to atoms in the BCC lattice point 
and deleted from the sample for better visualization of the collision cascade.
It is observed that the von Mises strain computation provides information of the 
geometry of the collision cascade, where atoms with a strain from 0.2 to 0.4 are found 
to be around the formed Frenkel pairs. 
These atoms are related to the counts maximum of the strain graph shown in 
Fig \ref{fig:fig3}d).
Therefore, the analysis of the damage in W material due to neutron bombardment can be 
done by quantifying Frenkel pair formed and identifying different defects geometries by 
FaVAD. 
In addition, the computation of atomic strains gives a visualization of the cluster size 
and geometry of the collision cascades.

\section{Concluding Remarks}
\label{conclusions}

In this paper, we performed machine learned based molecular dynamics simulations 
to emulate neutron bombardment on [111] W samples in a temperature range of 300 to 900K. 
Formation of Frenkel pairs are identified and quantified by the software workflow for 
fingerprinting and visualizing defects in damaged crystal structures (FaVAD), 
considering the magnitude of the thermal motion also. 
At room temperature, the number of Frenkel pair is 2 $\pm$ 1 as reported in the literature \cite{BJORKAS20093204,DOMINGUEZGUTIERREZ2020100724}. 
The increase of the sample temperature affects the atomic motion in the material and its
recovery after collision cascade, producing 4 $\pm$ 1 Frenkel pairs at the highest temperature. 
The computation of the atomic strain is used to visualize the geometry of atoms affected 
due to collision cascade and the cluster size, where W atoms with a strain between 0.2 and 0.4 
are found to be around the formed Frenkel pairs.

\section*{Acknowledgments}
We acknowledge support from the European Union Horizon 2020 research
 and innovation program under grant agreement no. 857470 and from the 
 European Regional Development Fund via the Foundation for Polish 
 Science International Research Agenda PLUS program grant 
 No. MAB PLUS/2018/8. We acknowledge the computational resources 
 provided by the Seawulf institutional cluster at the Institute for 
 Advanced Computational Science in Stony Brook University.


\bibliography{bibliography}
\bibliographystyle{iopart-num}
\end{document}